\documentclass[twocolumn,10pt]{IEEEtran}
\normalsize

\usepackage{soul}

\usepackage{enumerate}
\usepackage{color}

\usepackage{cite}
\usepackage{float}
\newfloat{figtab}{htb}{fgtb}
\makeatletter
\newcommand\figcaption{\def\@captype{figure}\caption}
\newcommand\tabcaption{\def\@captype{table}\caption}
\makeatother

\usepackage{graphicx}
\ifCLASSINFOpdf
\else
\fi

\usepackage{amsthm}
\usepackage{amssymb,amsfonts}
\usepackage{amsfonts,balance}

\usepackage[cmex10]{amsmath}

\newtheorem{theorem}{Theorem}

\usepackage{algorithmic}
\usepackage{array}
\usepackage{multirow,makecell}
\usepackage[caption=false,font=footnotesize]{subfig}

\usepackage{verbatim}

\newcommand{\db}[1]{{\textcolor{blue}{#1}}}
\newcommand{\todo}[1]{{\textcolor{black}{#1}}}

\begin{document}
\title{
CNN-Based Detection of Mixed-Molecule Concentrations in Molecular Communication}

\author{Vivien~Walter, Dadi~Bi,~\IEEEmembership{Member,~IEEE}, Daniel L. Ruiz Blanco, Yansha~Deng,~\IEEEmembership{Senior Member,~IEEE}\\
\thanks{Vivien Walter was with the Department of Engineering, King's College London, WC2R 2LS, U.K. He is now with i-magineXR, La Roche-sur-Yon 85000, France (e-mail: walter.vivien@gmail.com).}
\thanks{Dadi Bi was with the Department of Engineering, King's College London, WC2R 2LS, U.K. He is now with the James Watt School of Engineering, University of Glasgow, G12 8QQ, U.K. (e-mail: dadi.bi@glasgow.ac.uk). 
}
\thanks{Daniel L. Ruiz Blanco was with the Department of Engineering, King's College London, WC2R 2LS, U.K. He is now with Accenture, Madrid 28046, Spain (e-mail: d.ruiz.blanco@accenture.com).}
\thanks{Yansha Deng is with the Department of Engineering, King's College London, WC2R 2LS, U.K. (e-mail: yansha.deng@kcl.ac.uk). %(Corresponding author: Yansha Deng).
}
}

\maketitle

\begin{abstract}
Molecular communication (MC) is a promising paradigm for applications where traditional electromagnetic communications are impractical. However, decoding chemical signals, especially in multi-transmitter systems, remains a key challenge due to interference and complex propagation dynamics. In this paper, we develop a one-dimensional fractal convolutional neural network (fCNN) to detect the concentrations of multiple types of molecules based on the absorbance spectra measured at a receiver. Our model is trained by both experimental and simulated datasets, with the latter enhanced by noise modeling to mimic real-world measurements. We demonstrate that a noise-augmented simulated dataset can effectively be a substitute for experimental data, achieving similar decoding accuracy. Our approach successfully detects bit sequences in both binary and quadruple concentration shift keying (BCSK and QCSK) scenarios, even when transmitters are desynchronized, highlighting the potential of machine learning for robust MC signal detection.
\end{abstract}



% \fi
\IEEEpeerreviewmaketitle

\section{Introduction}

\label{sec_intro}

Molecular communication (MC) has emerged as a novel paradigm for transmitting information using chemical molecules, particularly in scenarios where traditional electromagnetic-based communication is unsafe or impractical, such as targeted drug delivery and explosive gas detection \cite{Akan2017,bi2021survey}. One of the primary research directions in MC is mathematical modeling of the propagation of signaling molecules and the derivation of the expected number of molecules observed at a receiver. To reduce analytical complexity, existing theoretical studies often make simplified assumptions, such as an unbounded propagation environment \cite{Pierobon2010,6708551}, perfect synchronization between transmitter and receiver \cite{6191345,6949028}, and ideal molecule counting at the receiver \cite{Jamali2019b}. Apart from the theoretical efforts, growing research has focused on experimental MC platforms that aimed at validating theoretical models and exploring real-world applications \cite{10105650,lotter2023experimental,10443866}.

%The recent years have witnessed a growing number of studies that applied machine learning to both theoretical and experimental MC research \cite{9247172} since some assumptions may not hold any more and considering the complexity of an experimental platform. For theoretical studies, it has been shown that how machine learning techniques can assist the channel modeling \cite{8277667,lee2017machine,9120354}, nanomachine or abnormality localization \cite{8964317,10748363}, and receiver design and signal detection \cite{8491088,kim2023machine,10059136,10130469,10904174}. For experimental studies, although experimental platforms can capture what molecules exactly experience and generate the most accurate response, there is still a need to develop a model to predict system outputs as experiments can be time-consuming, which would close the loop between theory and practice and enable to simulate, understand, and optimize system behavior across a wide range of conditions. 

%Recently, there has a growing interest in applying machine learning (ML) techniques to both theoretical and experimental MC research \cite{9247172}. This trend is driven by the limitations of oversimplified theoretical models, the violation of idealized assumptions in practice, and the inherent complexity of experimental systems.

With the progress from theoretical research to practical implementation, the limitations of oversimplified models and the violation of their assumptions become increasingly apparent. Although experimental MC systems are more realistic, they introduce additional layers of complexity that are often challenging to model analytically. These challenges have motivated the research that applied machine learning (ML) techniques to both theoretical and experimental MC research \cite{9247172}. In theoretical MC research, ML has been applied to assist with channel modeling \cite{8277667,lee2017machine,9120354}, localization of nanomachines or abnormalities \cite{8964317,10748363}, and the design of receivers and signal detection algorithms \cite{8491088,kim2023machine,10059136,10130469,10904174}.

%Initial works include the application of machine learning to the scenario when a single type of molecule is used as an information carrier, and it has been demonstrated that the vanilla recurrent neural network (RNN) can capture the channel and noise variation and detect the absence of glucose molecules propagated within a pipe \cite{9148818}. When multiple types of molecules are used to encode information within a pipe, it is also possible to employ support vector machines (SVM) and RNN-inspired algorithms to detect acids and bases \cite{8255058}, and principal component analysis (PCA) to classify different fluorescent quantum dots \cite{10443671}. Meanwhile, the feasibility of classification of different concentration levels of magnetic particles has been demonstrated with the use of either a conventional classifier with Linear Discriminant Analysis (LDA) or the newly constructed CNN (one dimensional CNN with nine layers) \cite{bartunik2022using}. Moreover, ML applied to more complex channel: and a long short-term with memory (LSTM) mode can accurately predict the pH value for the sciatic nerve communication channel of a bullfrog with collected compound action potentials (CAPs) \cite{deng2024channel}.

While experimental platforms can provide the most accurate insights into molecular behavior, they often require specialized expertise and can be time-consuming to operate. These limitations underscore the need for predictive models, which bridge theory and practice by enabling simulation, analysis, and optimization of system behavior under diverse operating conditions. Initial studies have focused on applying ML in experimental systems when a single type of molecule serves as the information carrier. For example, recurrent neural networks (RNNs) have been shown to effectively capture channel and noise variations and detect the presence or absence of glucose molecules propagating through a tubing channel \cite{9148818}. For systems employing multiple molecule types, support vector machines (SVMs) and RNN-based models have been employed to determine whether an acid or a base was transmitted \cite{8255058}, while principal component analysis (PCA) has been used to classify various fluorescent quantum dots \cite{10443671}. Additionally, classification of different concentration levels of magnetic particles has been demonstrated using both traditional classifiers based on linear discriminant analysis (LDA) and a custom-designed one-dimensional convolutional neural network (CNN) with nine layers \cite{bartunik2022using}. Beyond pipe-based propagation, ML has been applied to more complex channel scenarios. For instance, a long short-term memory (LSTM) network has been successfully used to predict pH levels in a sciatic nerve-based MC system in bullfrogs, based on collected compound action potential (CAP) signals \cite{deng2024channel}.

\begin{figure*}[!ht]
    \centering
    \subfloat[]{
    \includegraphics[width=.4\linewidth]{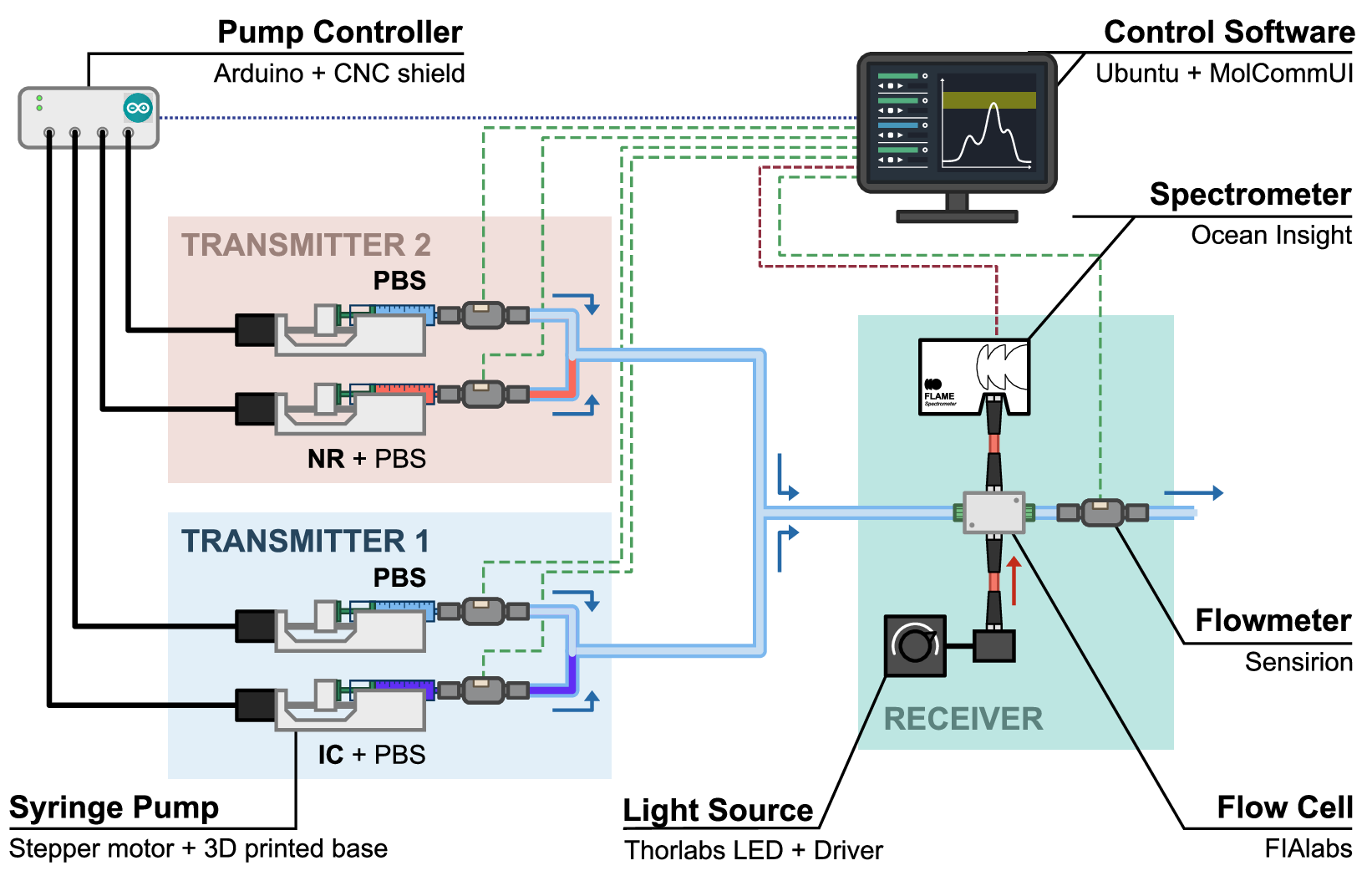}
    } 
    \subfloat[]{
    \includegraphics[width=.6\linewidth]{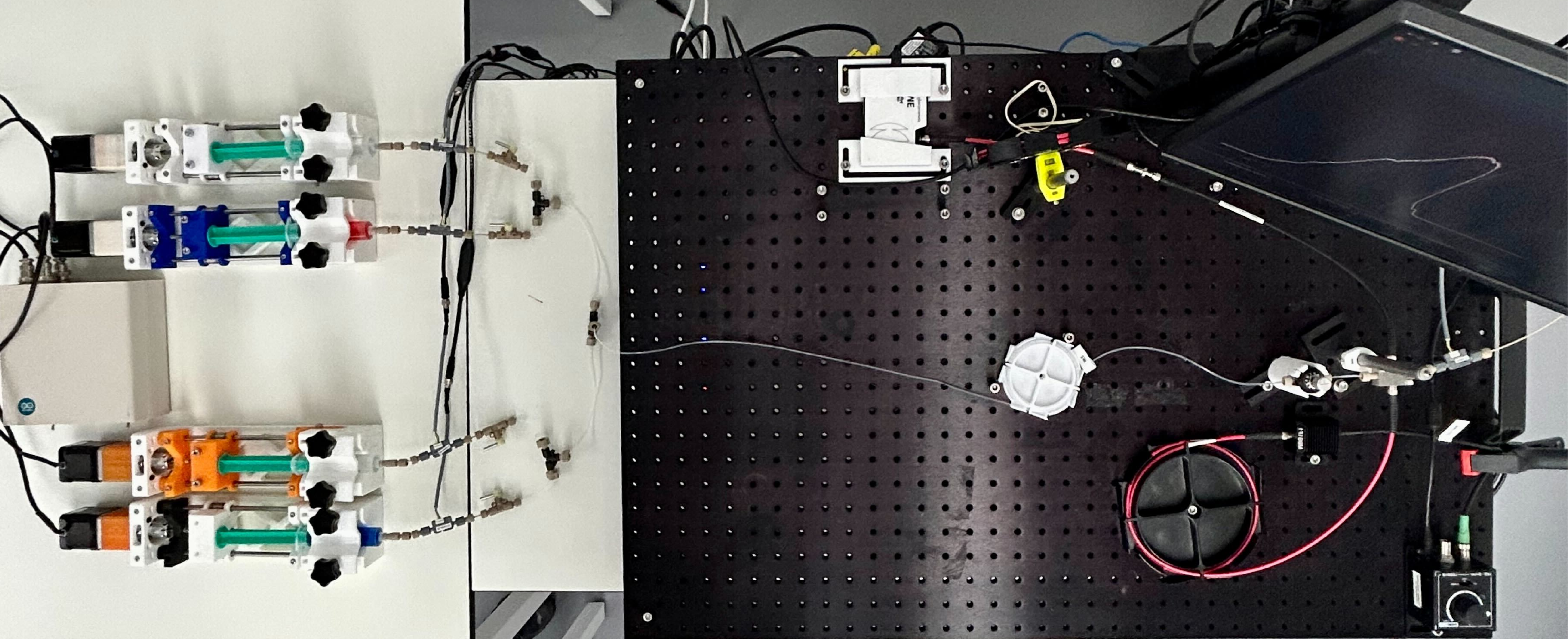}
    }
    \caption{Overview of the microfluidic molecular communication platform: (a) schematic illustration of the system architecture; (b) photograph of the assembled experimental setup.}
    \label{fig:testbed}
\end{figure*}

In this paper, we aim to propose a multi-molecule detection method for measuring the concentration values of different chemicals, rather than molecule type classification, on the microfluidic experimental platform as shown in Fig. \ref{fig:testbed}, where multiple types of dyes emitted from distinct transmitters propagate through tubing and are detected by an ultraviolet-visible (UV-Vis) spectrometer. This platform is inspired by our microfluidic molecular communication (MIMIC) experimental setup presented in \cite{walter2023real}, which validated the feasibility of chemical signal processing for MC using the universally known acid and base-based chemical reactions. In MIMIC, acid or base solutions emitted by a single transmitter interact with a pH-sensitive dye and are detected by the same UV-Vis spectrometer. Thus, the current platform can be viewed both as a simplified or adapted version of MIMIC and as a generalizable setup applicable to other experimental configurations, such as molecular shift keying (MoSK) modulation and multiple-input single-output (MISO) systems.

Deriving a closed-form expression for the concentration of different types of transmitted molecules in this setup is highly challenging due to the parabolic flow dynamics and the intricately structured internal geometry of the system components. In addition, two key communication challenges further complicate the system. First, the presence of multiple transmitters and varying concentration levels for signal encoding leads to a combinatorial explosion in the number of signals to be detected by the receiver.
%Using concentration shift keying (CSK), each transmitter encodes information as a specific concentration of its respective molecule, and signal overlap among the M transmitters results in n^M possible combinations that must be distinguished at the receiver. 
Second, when multiple transmitters are not synchronized with each other and emit signals at different times, they can cause temporal variations in the chemical environment, thereby complicating signal detection. To tackle these challenges, we make the following contributions:
\begin{itemize}
    \item We propose a one-dimensional fractal convolutional neural network to directly detect the concentrations of different types of chemical molecules from UV-Vis absorbance measurements. Unlike prior works  \cite{9148818,8255058,10443671,bartunik2022using}, which focused on classification, our method enables continuous concentration detection for high-resolution signal recovery.      
    \item We address the data scarcity challenge in experimental MC by proposing a training strategy based on simulated spectra. These synthetic datasets, generated using the Beer–Lambert law and augmented with noise model fitted by experimental data, allow for efficient and scalable training without relying on labor-intensive lab measurements.    
    \item We demonstrate that our proposed CNN trained on simulated data with noise achieves detection accuracy comparable to models trained on experimental data. Our model successfully decodes binary and quadruple concentration shift keying (BCSK and QCSK) signals—including asynchronous transmissions—validating its practical utility in real-world chemical communication scenarios.
\end{itemize}

The rest of this paper is organized as follows. In Sec. \ref{sec_testbed}, we introduce our experimental MC platform. The CNN framework along with the preparation of the training data are provided in Sec. \ref{sec_model}, and the results that validate the CNN framework are illustrated in Sec. \ref{sec_result}. Finally, we conclude the paper in Sec. \ref{sec_conclusion}.

\section{Experimental Platform}
\label{sec_testbed}

To demonstrate how a CNN can be used to address the communication challenges mentioned in Sec. \ref{sec_intro}, we built the experimental platform illustrated in Fig.~\ref{fig:testbed} that consists of two transmitters and one receiver. It is noted that we used two transmitters to demonstrate the feasibility of our proposed methodology introduced in Sec. \ref{sec_model}, but the platform and methodology can be extended to multiple transmitters. For simplicity, we refer to these two transmitters as ${\rm TX}_1$ and ${\rm TX}_2$. In this experimental setup, each transmitter consists of two 3D-printed syringe pumps connected to a single outlet via a T-shaped junction (OD 1/16 inch, ID 0.050 inch, IDEX Material Processing Technologies, USA). The first syringe pump is filled with a colored solution stabilized at pH~7.4 using phosphate buffer saline (PBS), which will encode the message; the second pump is filled with the corresponding solvent, i.e., a PBS solution at pH~7.4, which is used to maintain a constant flow rate at the transmitter output. We selected Indigo Carmine (IC) and Neutral Red (NR) to prepare the colored solutions for ${\rm TX}_1$ and ${\rm TX}_2$ to transmit signals and train the CNN, with all chemicals purchased from Merck (ex-Sigma Aldrich, Germany).

The transmitters are connected to the inlets of a T-shaped junction, with the other side connecting to the inlet of a single receiver. Therefore, the microfluidic tubing between one end of the T-shaped junction and the receiver inlet can be considered as the propagation channel. The receiver consists of a flow cell (Z-type 2.5mm pathlength, stainless steel, FIAlabs instruments, USA) connected to an LED light source (470–850nm fibre-coupled LED, Thorlabs, USA) and an Ultraviolet-Visible spectrometer (FLAME-T-UV-Vis spectrometer, 3648px, Ocean Insight, USA) which continuously read the absorbance spectrum of the solution going through the flow cell in real-time. {The selected UV-Vis spectrometer can read 3648 different wavelengths simultaneously, and therefore it can detect and identify a larger number of signaling molecules compared to a pH meter \cite{9148818} and a susceptibility sensor \cite{bartunik2022using}. } The transmitters, propagation channel, and receiver were all assembled using the microfluidic tubing with OD 1/16 inch and ID 0.75mm.

\begin{figure}[!t]
    \centering
    \includegraphics[width=\linewidth]{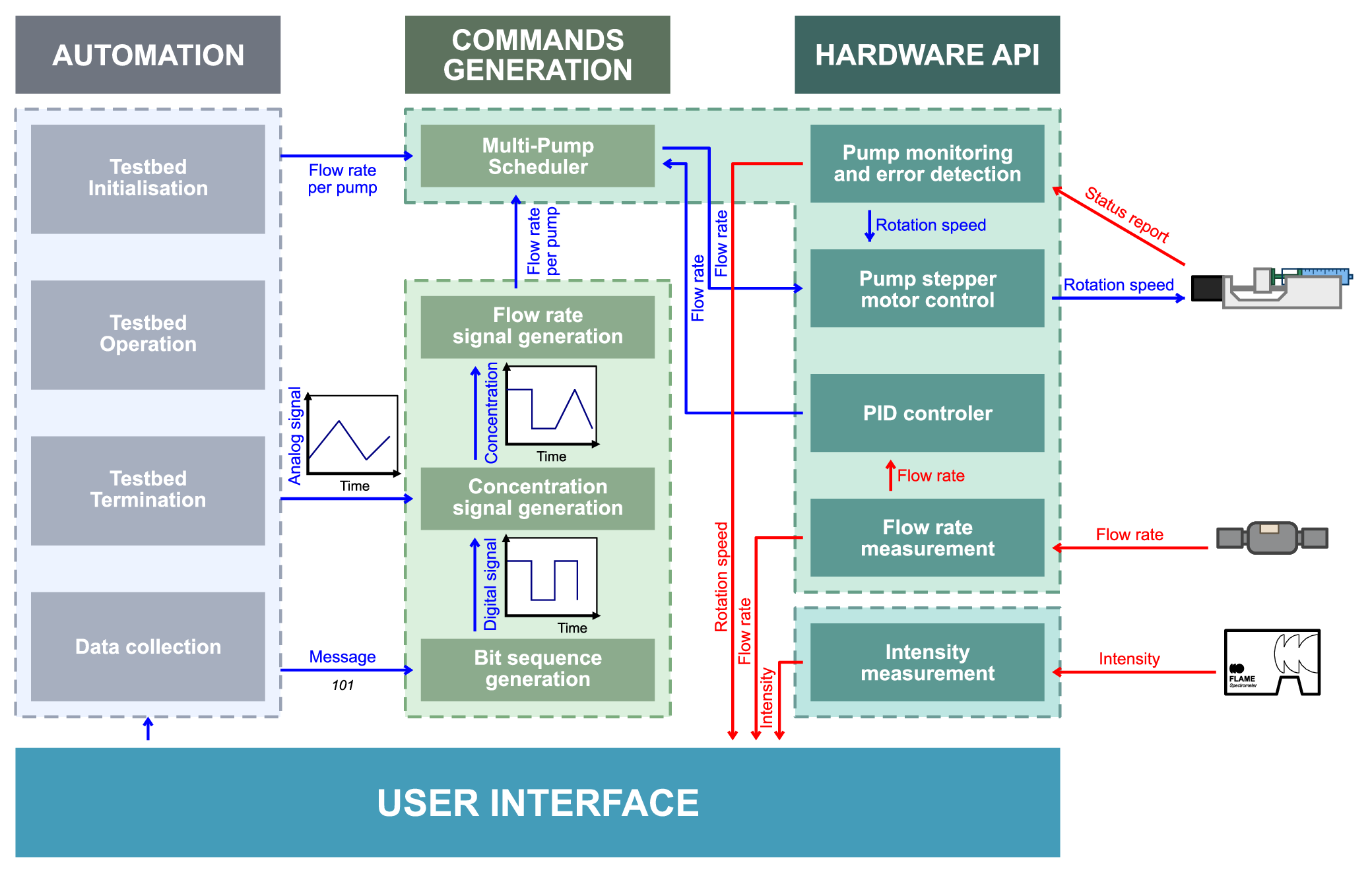}
    \caption{Architecture of the developed MolCommUI software.}
    \label{fig:software}
\end{figure}

Throughout the experiments, the transmitters and the receiver are operated and monitored by a developed software called MolCommUI \cite{walter2023real}, which is implemented using Python and its architecture is illustrated in Fig. \ref{fig:software}. Briefly, MolCommUI processes input signals by translating them into a series of commands that are transmitted to an Arduino, which regulates the flow rate of pumps while simultaneously collecting data from flow meters and the spectrometer. To enhance the stability of the pump flow rate during injection, a feedback control mechanism was implemented, dynamically adjusting the pump motor speed using a tuned proportional–integral–derivative (PID) algorithm.

%This requirement is satisfied in our experimental setup through the use of a UV-Visible (UV-Vis) spectrometer as the sensor of the receiver, which can read 3648 different wavelength simultaneously (INSERT MAKE AND MODEL).

\section{CNN for Concentration Regression}
\label{sec_model}

In this section, we first provide the mathematical expression of the absorbance spectra observed by the UV-Vis spectrometer at the receiver. We then introduce the architecture of the CNN framework and finally present the training and testing steps. 

\subsection{Characterization of Absorbance Spectra}
Although there are only two transmitters considered in our platform (see Fig. \ref{fig:testbed}), here we provide a general method to analyze the quantitative data collected by a single receiver and extract the messages emitted by $M$ transmitters at the same time. For our MC platform, transmitted signals will be extracted from the absorbance spectra collected by the selected UV-Vis spectrometer. In the presence of a mixture of $M$ different colored molecules, the absorbance $A$ of the solution measured by the receiver at time $t$ and at the wavelength $\lambda$ can be approximated by the Beer-Lambert law as \cite{walters1997spectrophotometric} 
\begin{equation}
    \label{eq:beer_lambert}
    A\left( \lambda, t \right) = l \sum^{M}_{i} \left( \varepsilon_i (\lambda) C_i (t) \right),
\end{equation}
with $l$ as the distance travelled by the light through the solution (i.e., the length of the flow cell used in our receiver), $\varepsilon_i (\lambda)$ as the molar extinction coefficient of the molecule $i$ at the wavelength $\lambda$ (i.e., a measure of how strongly a chemical species absorbs light at a given wavelength), and $C_i(t)$ as the concentration of the molecule $i$ at time $t$. %In our experimental setup, the UV-Vis spectrometer can read a total of 3648 different wavelengths at each measurement. %{\color{red}\st{In order to decode the messages sent by the $N$ transmitters, the receiver must extract from the absorbance measurement all the concentrations $C_i(t)$ for all the $N$ molecules.}}

\subsection{Fractal CNN Architecture}

%Here, we suggest a regression of the absorbance spectrum to extract the concentrations of $N$ molecules through the use of a 1-D fractal Convoluted Neural Network (fCNN). This approach is based on the architecture proposed by \cite{larsson_2017} and has been successfully applied to UV-Vis spectroscopy for quantifying dissolved gases in turbid solutions by \cite{xia_2023}.  

%\db{To extract the concentrations of $N$ molecules, here we suggest a regression of the absorbance spectrum through the use of a 1-D fractal Convoluted Neural Network (fCNN). The motivation of using this architecture is that CNN exhibits a better performance for spectral data (spectrometric quantification) compared to methods like partial least squares regression (PLSR) [], principal component regression (PCR) [], and support vector machine (SVM) [], and fractal CNN allows very deep networks without an explosion in parameters has been successfully applied to UV-Vis spectroscopy for quantifying dissolved gases in turbid solutions by \cite{xia_2023}. }

\begin{figure}[!t]
\centering
\includegraphics[width=\linewidth]{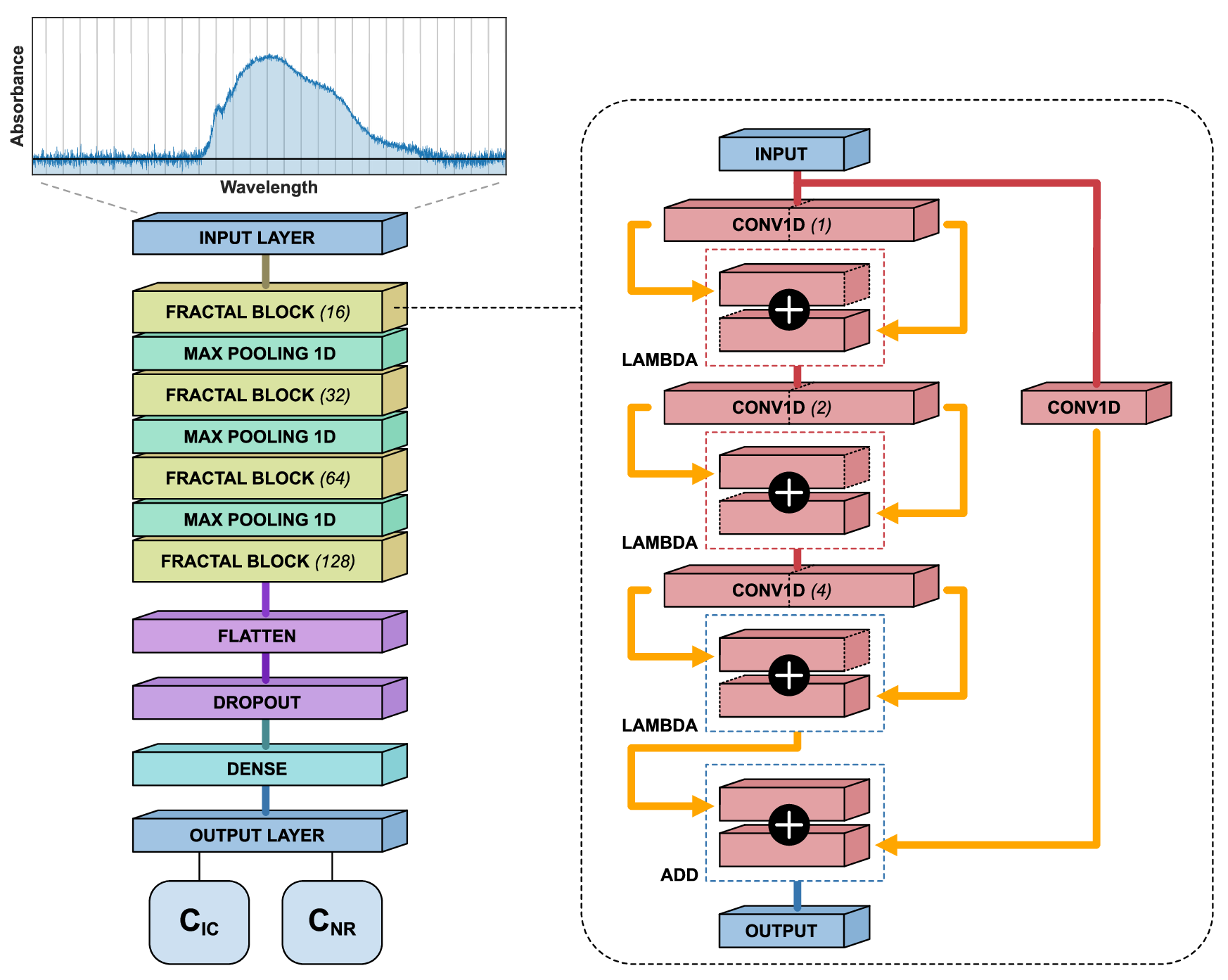}
\caption{Architecture of the 1-D fractal convoluted neural network (fCNN).}
\label{fig:fcnn_architecture}
\end{figure}

To detect the concentrations of $M$ molecular species from the received mixed solution, we propose a one-dimensional fractal convolutional neural network (fCNN) applied to the absorbance spectrum, as shown in Fig. \ref{fig:fcnn_architecture}. This architecture is motivated by the superior performance of CNNs in spectrometric quantification compared to conventional techniques such as partial least squares regression (PLSR) \cite{langergraber2003multivariate}, principal component regression (PCR) \cite{guan2018research}, and SVM \cite{lu2021detection,wolf2013predicting}. Moreover, the fCNN structure enables the construction of very deep networks without a significant increase in the number of parameters \cite{larsson_2017}, and it has been successfully employed in UV-Vis spectroscopy for quantifying dissolved gases in turbid media, as demonstrated in \cite{xia_2023}. As illustrated in Fig.~\ref{fig:fcnn_architecture}, the 1D input vector, comprising the absorbance spectrum of the received solution measured at 3648 discrete wavelengths, is fed into a sequence of four feature extraction layers, referred to as ``fractal blocks”. These blocks are interleaved with max pooling layers to progressively reduce the dimensionality and extract hierarchical features. The output from the final fractal block is passed to a regression module, which consists of a flatten layer, a dropout layer with a dropout rate of 0.5 to mitigate overfitting, and a dense output layer with linear activation. The output dimension of the fCNN corresponds to the number of molecular species to be identified, which is two in the illustrative example shown in Fig.~\ref{fig:fcnn_architecture}.

%The four fractal blocks of the fCNN are characterized by their increasing filter size, from 16 to 128. Each block is made up of three 1-D convolutional layers, {\color{red}with twice the input filter size of the block?}, a kernel size of 3, a ReLu activation and increasing dilation rates from 1 to 4. Each 1-D convolutional layer is followed by a Lambda layer, where the first half of the output vectors is summed with the second half. The final layer of the block is an Add layer, which sums the outputs of the three consecutive 1-D convolutional layers with the output of an additional 1-D convolutional layer applied directly to the input vector of the fractal block. This additional convolutional layer uses the same filter size as the block and a kernel size of 1. All 1-D convolutional layers used in the fractal block use an even padding of the input.

Each of the four fractal blocks in the fCNN architecture is defined by an increasing number of filters, ranging from 16 to 128. Within each block, three consecutive one-dimensional convolutional layers (CONV1D) are used, each with a filter size equal to twice the value of the fractal block, a kernel size of 3, ReLU activation, and dilation rates increasing from 1 to 4. Following each convolutional layer, a Lambda layer is employed to sum the first and second halves of the output vector produced by a CONV1D. The outputs of the three convolutional layers are then combined with the output of a shortcut 1-D convolutional layer—applied directly to the input of the block—via an Add layer. This shortcut layer uses the same number of filters as the block and a kernel size of 1. All convolutional operations within the fractal blocks employ ``same'' padding to maintain the input length.

The performance of the fCNN model is evaluated using the mean square error (MSE) function, and the accuracy of the detection is quantified using the coefficient of determination $D$ defined in \cite{xia_2023} for a similar fCNN architecture as
\begin{equation}
    \label{eq:coeff_determination}
    D = 1-\frac{ \sum\limits^M_{i=1} \left( C_i - C_i' \right)^2 }{\theta + \sum\limits^M_{i=1} \left( C_i - \bar{C} \right)^2},
\end{equation}
where $C_i'$ is the detected concentration for the $i$th molecule, $C_i$ is the corresponding actual concentration value, $\bar{C}$ is the mean of all real concentration values, and $\theta$ is a small constant to prevent division by zero when all detected concentrations equal the real concentrations.

\begin{figure}[!t]
	\centering	\includegraphics[width=\linewidth]{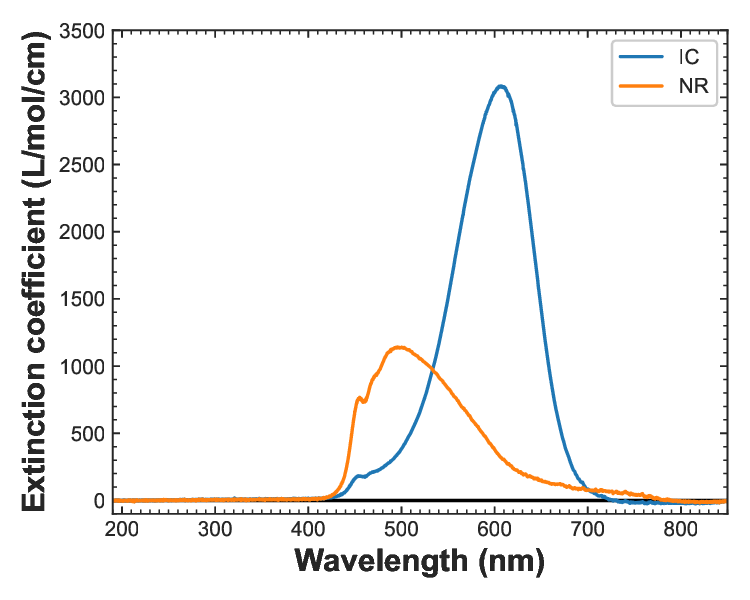}
	\caption{The extinction coefficient spectrum of Indigo Carmine (IC) and Neutral Red (NR).}
	\label{fig_absorbance}
\end{figure}
\subsection{Spectroscopy Datasets} 
In this paper, we selected Indigo Carmine (IC) and Neutral Red (NR) as the information carriers for ${\rm TX}_1$ and ${\rm TX}_2$ to transmit signals. As shown in Fig. \ref{fig_absorbance}, these two pH indicators exhibit different maximum peak extinction coefficient\footnote{To highlight the intrinsic chemical properties of IC and NR rather than experimental conditions, the Y-axis of Fig. \ref{fig_absorbance} represents the extinction coefficient instead of absorbance. Unlike the extinction coefficient, absorbance is a strict \textit{in situ} measurement that depends on numerous factors, including the quality and geometry of the experimental setup, as well as the purity and composition of the solvent. Consequently, absorbance is not considered a direct chemical property of the molecule.} values at pH 7.4, with peaks at 608 nm and 496 nm, respectively \cite{sabnis_book}. In the following, we introduce two types of training datasets for our proposed fCNN: 1) a pure experimental dataset, 2) a simulation dataset.

%The source can be found from: /Users/k1898193/Library/CloudStorage/OneDrive-King'sCollegeLondon/Microfluidic MC Projects/Second Paper/sources/Experiments/2. Ph dyes/Dye analysis

\subsubsection{Experimental Dataset}
\begin{figure}[!t]
	\centering
	\includegraphics[width=\linewidth]{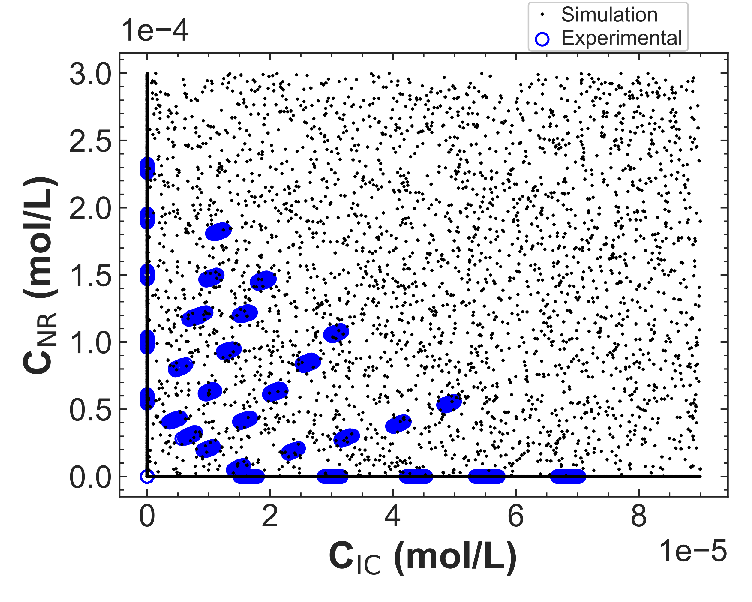}
	\caption{The distribution of Indigo Carmine (IC) and Neutral Red (NR) concentrations across samples prepared either experimentally or through simulation.}
	\label{fig_training_dataset}
\end{figure}

The experimental dataset was generated by dissolving IC and NR in PBS at pH 7.4, followed by the preparation of multiple solutions with varying concentrations of IC and NR. The concentration of IC and NR ranged from 0 mol/L to the maximum concentration permitted by the transmittance of each chemical, defined as the percentage of light passing through a colored solution compared with a transparent one, approximately $7 \times 10^{-5}$ for IC and $2.5 \times 10^{-4}$ for NR). A total of 31 different IC/NR mixtures were prepared and classified into four groups: 1) IC only, (2) NR only, (3) both IC and NR, and (4) neither IC nor NR (i.e., pure PBS). This resulted in 12000 data points, excluding the pure PBS samples. Each solution was manually injected into a flow cell connected to a spectrometer, and its absorbance was measured. We plot the concentration distribution of these samples using blue color in Fig. \ref{fig_training_dataset}.

\subsubsection{Simulation Dataset}
Considering that the obtained experimental data is limited in the number of different concentration combinations of IC and NR due to time, cost, or practical constraints in laboratory procedures, a simulated dataset was also prepared to serve as a complete and independent training dataset. The simulated dataset allows for the rapid generation of a wider range of concentration combinations that may not be easy to capture experimentally, and helps improve the robustness and generalization capability of the fCNN model. Following the process described in the following paragraphs, we prepared a total of 12000 data points to match the number collected experimentally. We also explored higher concentrations beyond the experimentally achievable range to improve the detection accuracy of the fCNN model. A comparison of the concentration distributions of these 12,000 data points and the experimental dataset is shown in Fig. \ref{fig_training_dataset}.

%The generation of a simulated spectroscopic dataset is based on the Beer-Lambert law provided in \eqref{eq:beer_lambert}: if the molar extinction coefficient profile over the wavelengths $\varepsilon(\lambda)$ of each molecule is known, the absorbance spectrum of a solution can be approximated by the sum of these profiles multiplied by their respective concentrations. The extraction of the $\varepsilon(\lambda)$ profile for each molecules was achieved by using similar solutions as the ones used to collect the experimental training dataset and measuring their absorbance using the spectrometer. Because of potential energetic interactions between the two molecules in solution, we included in this profile extractions dataset solutions with mixtures of both molecules. The dataset were then fitted 3648 times, once for each different wavelength, using the equation

The simulated spectroscopic dataset was generated based on the Beer–Lambert law in \eqref{eq:beer_lambert}, which requires the molar extinction coefficient profiles across different wavelengths $\varepsilon(\lambda)$ for IC and NR. The values of $\varepsilon(\lambda)$ were experimentally determined (see Fig. \ref{fig_absorbance}) using solutions prepared in a manner consistent with those used to construct the experimental training dataset. To account for potential energetic interactions between IC and NR in mixed solutions, $\varepsilon(\lambda)$ was fitted independently at 3648 distinct wavelengths based on measured absorbance data using
\begin{equation}
    \label{eq:fit_equation}
    \frac{A}{l} = \varepsilon_{\rm IC} C_{\rm IC} + \varepsilon_{\rm NR} C_{\rm NR},
\end{equation}
where $l$ is the length of the flow cell and is fixed at 0.25~cm throughout this paper. The resulting extinction coefficient profiles, $\varepsilon_{\rm IC}(\lambda)$ and $\varepsilon_{\rm NR}(\lambda)$, were then employed to generate the simulated dataset according to \eqref{eq:beer_lambert}. We also plot the distribution of the simulated concentrations using black dots in Fig.~\ref{fig_training_dataset}.

%However, the process of extracting molar extinction coefficients for the generation of simulated data inherently removes sensor-related noise present in the experimental measurements. 

%While the molar extinction coefficient enables the construction of a simulated dataset, it is an intrinsic and noise-free chemical property. As a result, the simulated absorption spectra calculated via \eqref{eq:fit_equation} are free from sensor-related noise and appear significantly cleaner than their experimental counterparts, potentially leading to discrepancies in model training and generalization. 

While the molar extinction coefficient allows the construction of a simulated dataset, it is an intrinsic and noise-free chemical property, and therefore the simulated absorption spectra calculated via \eqref{eq:fit_equation} are free from sensor-related noise. The experimental dataset can also be subject to additional sources of noise during sample preparation, that introduced by pipetting or weighing. As a result, the simulated dataset appears significantly cleaner than its experimental counterparts, potentially leading to discrepancies in model training and generalization.

%To mitigate the impact of this discrepancy, we incorporated sensor noise into the simulated data. This was achieved by first characterizing the noise profile from the experimental dataset and then synthetically generating random noise to be added to the simulated spectra. 

We incorporated sensor noise into the simulated data by first characterizing the noise profile from the experimental dataset and then synthetically generating random noise to be added to the simulated spectra. Assuming the sensor noise is unbiased and randomly distributed, it can be modeled as a Gaussian distribution, i.e., $n \sim \mathcal{N}(\mu, \sigma^2)$, with zero mean and a standard deviation proportional to the measured light intensity. Based on the experimental dataset, we found that at low light intensity $I_{\text{Min}}$, the sensor noise can reach up to $e_\text{Max}=2\%$ of the intensity, whereas at high light intensity $I_{\text{Max}}$, the noise decreases to as low as $e_\text{Min}=0.5\% $. Between these extremes, the noise amplitude appears to decrease approximately linearly with increasing light intensity. Here, the values of $I_{\text{Min}}$ and $I_{\text{Max}}$ are specific to the UV-Vis spectrometer in use. Given the absorbance $A$, calculated from \eqref{eq:fit_equation}, and a constant light source with incident intensity $I_0$, the transmitted light intensity $I$ is obtained as
\begin{equation}
    I=\frac{I_0}{10^A}.
\end{equation}
The transmitted intensity can then be normalized as
\begin{equation}
    I_{\text{Norm}}=\frac{I-I_{\text{Min}}}{I_{\text{Max}}-I_{\text{Min}}}.
\end{equation}
With the noise amplitude varying linearly with the normalized intensity, the variance of the noise distribution is given by 
\begin{equation}
    \sigma^2=I_{\text{Norm}} (e_\text{Max}-e_\text{Min})+e_\text{Max}
\end{equation}
Finally, incorporating Gaussian noise into the transmitted intensity, the resulting absorbance with noise is expressed as
\begin{equation}
    A_\text{noise}=\log_{10}\frac{I_0}{I(1+n)}.
\end{equation}

\begin{figure}[!t]
    \centering
    \includegraphics[width=\linewidth]{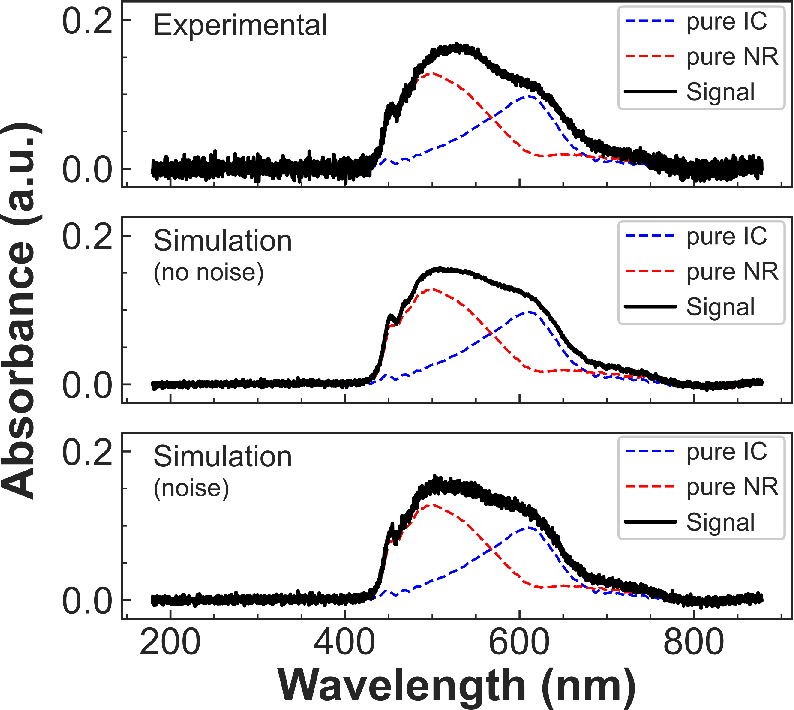}
    \caption{Comparison of the absorbance spectrum of the same sample (i.e., same $\rm C_{IC}$ and $\rm C_{NR}$) obtained through experiment and simulation.}
    \label{fig_signal_comparison}
\end{figure}

Comparisons among the experimental spectra, noise-free simulated spectra, and noise-augmented simulated spectra are presented in Fig.~\ref{fig_signal_comparison}. The importance of including sensor noise was further evaluated by training the fCNN model on simulated data both with and without the addition of noise.

\begin{comment}
    \begin{figure*}[!t]
    \centering
    \includegraphics[width=\linewidth]{figures/training.eps}
    \caption{Training of the fCNN on different dataset. (A) Distribution in concentration in Indigo Carmine (IC) and Neutral Red (NR) of the different sample prepared either experimentally or by simulation. (B) Comparison of the absorbance spectrum of the same sample (same $C_{\rm IC}$ and same $C_{\rm NR}$) obtained experimentally, by simulation without noise or by simulation including noise. The absorbance profile of the pure chemicals are presented in dashed lines. (C) Accuracy of the different prediction models on the concentration of each molecule. The blue dashed line represents the expected result of a perfect prediction, and the pink area indicates the concentration range which is too low to be analyzed.}
    \label{fig:fcnn_training}
    \end{figure*}
\end{comment}

\section{Results and Analysis}
\label{sec_result}

In this section, we evaluate the performance of the fCNN model proposed in Sec. III. In particular, three fCNN models are implemented in Python 3.10 using TensorFlow 2.12 and were trained separately using 1) the experimental dataset, 2) the simulated dataset without sensor noise, and 3) the simulated dataset augmented with sensor noise. According to \cite{xia_2023}, the training of each fCNN was divided into three consecutive optimization phases. Each phase was run on the same training dataset and, when applicable, initialized with the optimal parameters obtained from the previous phase using TensorFlow’s model checkpoint system. All three phases employed the Adam optimizer but differed in their learning rates: 0.001 for the first phase, 0.0001 for the second, and 0.00001 for the third and final phase. The remaining parameters of the Adam optimizer were kept consistent across phases: 200 epochs per phase (for a total of 600 epochs on all phases combined), 100 steps per epoch, batch size of 10, $\beta_1 = 0.9$, $\beta_2 = 0.999$, and $\epsilon = 1 \times 10^{-8}$. The small constant in \eqref{eq:coeff_determination} is set as $\theta=1 \times 10^{-7}$. For the artificial noise, $I_{\text{Min}}=3000$ and $I_{\text{Max}}=45000$. To compare the different fCNN models, the original dataset is shuffled and split into a training set (80\%) and a validation set (20\%).

\subsection{Performance of the fCNN Model}

\begin{figure}[!t]
    \centering
    \includegraphics[width=0.65\linewidth]{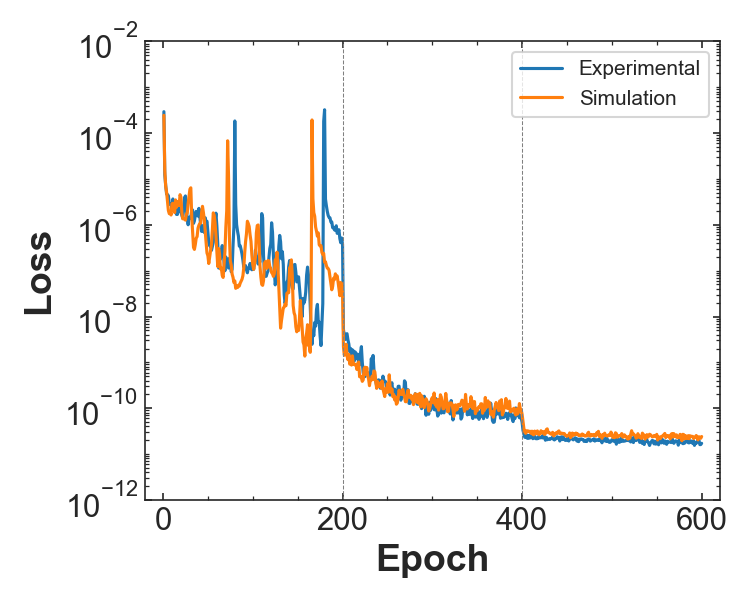}
    \caption{Convergence trend of the models trained by the experimental dataset and the simulated dataset with added sensor noise.}
    \label{fig_loss}
\end{figure}

\begin{table*}[!t]
\caption{Performance of the different fCNN models, evaluated on the same validation dataset. All values are given in mol/L. \label{table:model_accuracy}}
\centering
\begin{tabular}{|l||c|c|c|c|}
\hline \multirow{2}{*}{\textbf{Training Dataset}}
 & \multicolumn{2}{c|}{\textbf{Indigo Carmine (IC)}} & \multicolumn{2}{c|}{\textbf{Neutral Red (NR)}} \\
\cline{2-5}
 & \textbf{Min. Concentration} & \textbf{Detection Error} & \textbf{Min. Concentration} & \textbf{Detection Error} \\
\hline
Experimental & $3.7 \times 10^{-6}$ & $1.8 \times 10^{-6}$ & $1.2 \times 10^{-5}$ & $2.5 \times 10^{-6}$ \\
\hline
Simulation (without noise) & $4.3 \times 10^{-5}$ & $6.2 \times 10^{-6}$ & $8.3 \times 10^{-5}$ & $1.5 \times 10^{-5}$ \\
\hline
Simulation (with noise) & $1.1 \times 10^{-5}$ & $2.2 \times 10^{-6}$ & $2.7 \times 10^{-5}$ & $2.5 \times 10^{-6}$ \\
\hline
\end{tabular}
\end{table*}
\begin{figure}[!t]
    \centering
    \includegraphics[width=\linewidth]{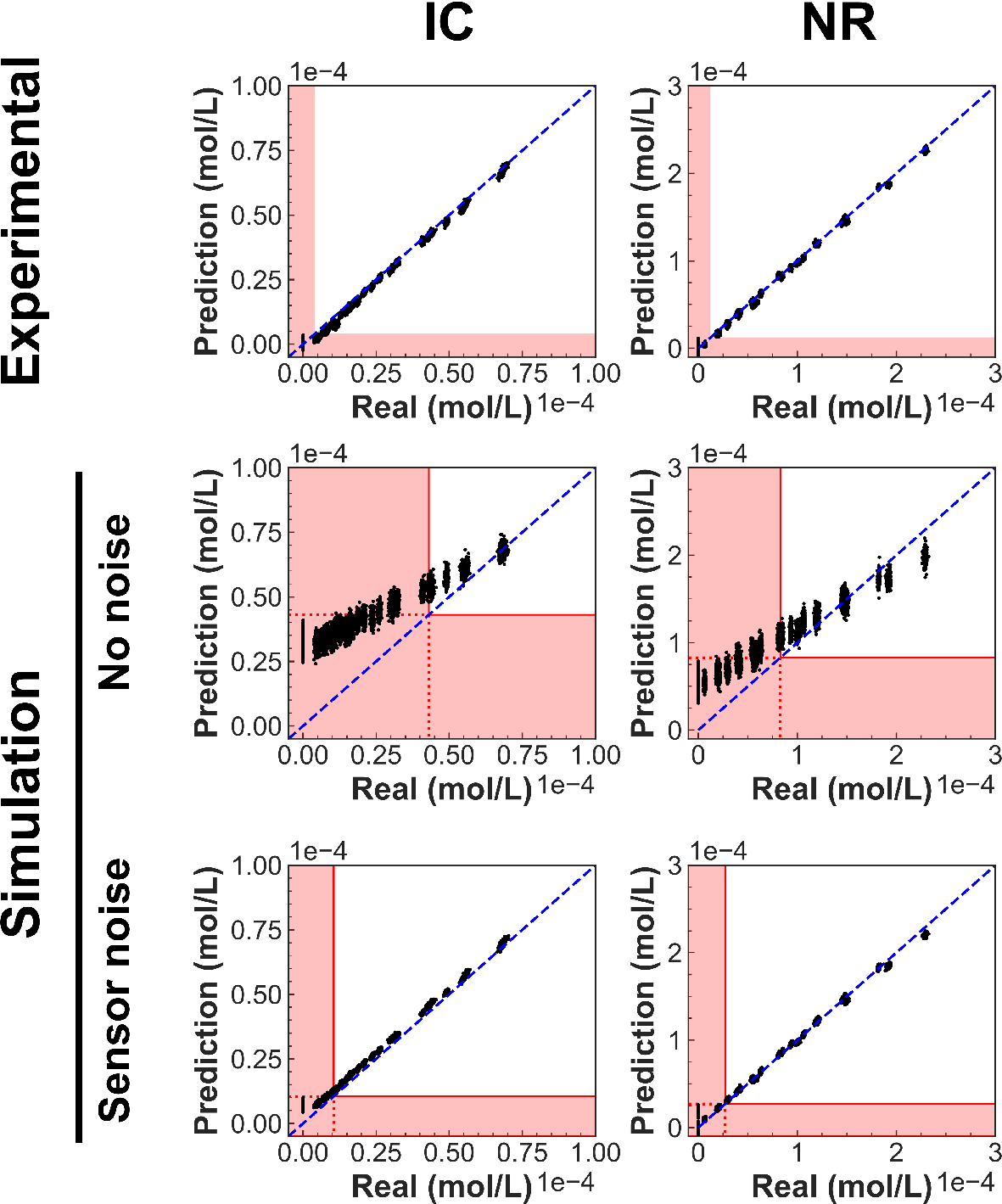}
    \caption{Accuracy of the different detection models on the concentration of each molecule. The blue dashed line represents the expected result of a perfect detection, and the pink area indicates the concentration range which is too low to be analyzed.}
    \label{fig_impact_of_training}
\end{figure}
To investigate the impact of different training datasets, the performance of the three trained fCNN models was evaluated using the same validation dataset. In particular, we focused on two metrics: 1) the minimum detectable concentration, defined as the highest predicted concentration among samples with a real concentration of zero (originated from prediction/statistical noises generated by the model), and 2) the detection error for concentrations above the minimum detectable concentration. The minimum detectable concentration was determined by inputting the absorbance spectrum of pure water into the fCNN and recording its maximum detected value, whereas the detection error was quantified as the average root mean square error (RMSE) between the detected values and the real experimental values, restricted to concentrations above the minimum detectable concentration. The results of these two metrics are summarized in Table \ref{table:model_accuracy}, and the comparison of the predicted concentration and actual experimental value is plotted in Fig. \ref{fig_impact_of_training}. From Table \ref{table:model_accuracy} and Fig. \ref{fig_impact_of_training}, it shows that our proposed fCNN model trained on the experimental dataset and the simulated dataset with sensor noise can accurately detect the concentration of experimental validation samples with comparable performance. Specifically, the minimum detectable concentration of the noise-augmented simulated model is less than three times higher than that of the experimental model, {while the detection error remains nearly identical—1.2 times higher for IC and equivalent for NR}. Moreover, as shown in Fig. \ref{fig_loss}, the training across these two datasets exhibits similar convergence behavior within the same time frame.

By contrast, the fCNN model trained on the simulated dataset without sensor noise demonstrates a worse performance. From Table \ref{table:model_accuracy}, the minimum detectable concentration of the fCNN model trained on the simulated dataset without sensor noise is up to ten times higher than that of the experimentally trained model for IC, and its detection error increased by as much as sixfold for NR. As shown in Fig. \ref{fig_impact_of_training}, the detection does not align with the reference blue line, and the detection error arises from inaccurate estimations across the entire concentration range, where it overestimates low concentrations and underestimates high concentrations. Importantly, once sensor noise was incorporated into the simulated training dataset, the accuracy of the fCNN model can be improved substantially, achieving performance nearly equivalent to that of the model trained on experimental data. These findings confirm that a simulated dataset of the same size as an experimental dataset can be effectively used to train an fCNN model for accurate concentration detection when it is appropriately augmented with sensor noise, resulting in only a marginal increase in detection error.

\begin{comment}
\begin{table*}[!t]
\caption{Efficiency of the different fCNN models, evaluated on the same dataset of experimental samples. All values are given in mol/L. \label{table:model_accuracy}}
\centering
\begin{tabular}{|l||c|c|c|c|}
\hline 
 & \multicolumn{2}{c|}{\textbf{Indigo Carmine (IC)}} & \multicolumn{2}{c|}{\textbf{Neutral Red (NR)}} \\
\hline
\textbf{Training Dataset} & \textbf{Min. Concentration} & \textbf{Prediction Error} & \textbf{Min. Concentration} & \textbf{Prediction Error} \\
\hline
Experimental & $3.7 \times 10^{-6}$ & $1.8 \times 10^{-6}$ & $1.2 \times 10^{-5}$ & $2.5 \times 10^{-6}$ \\
\hline
Simulation (without noise) & $4.3 \times 10^{-5}$ & $6.2 \times 10^{-6}$ & $8.3 \times 10^{-5}$ & $1.5 \times 10^{-5}$ \\
\hline
Simulation (with noise) & $1.1 \times 10^{-5}$ & $2.2 \times 10^{-6}$ & $2.7 \times 10^{-5}$ & $2.5 \times 10^{-6}$ \\
\hline
\end{tabular}
\end{table*}
\end{comment}

\subsection{Application to Communication}

With the validated fCNN models trained by the experimental and simulated datasets, we then applied them to demodulate the messages emitted by the two independent transmitters. As illustrated in Fig. \ref{fig:testbed}, $\rm TX_1$ modulates bits into the concentration of IC, while $\rm TX_2$ modulates bits into the concentration of NR. The performance of the fCNN models was assessed by two modulation schemes: binary concentration shift keying (BCSK) and quadruple concentration shift keying (QCSK). 
%{\color{red}For both the modulation methods, the communication was divided into intervals with the same duration $T_b=?$ during which the transmitter was scheduled to convey one-bit information per interval. For binary bit-1, the syringe pumps were turned on to inject chemical signal Y and solution P from the beginning of the bit interval until time $\alpha T_b$, with $\alpha$ as the duty cycle ($\alpha \leq 1$), and were turned off for the remaining time of the bit interval acting as a guard interval. For binary bit-0, the syringe pumps containing Y and P were turned off during the whole bit interval. }

\subsubsection{BCSK}
In BCSK, the signal was converted at each transmitter by modulating bit-1 onto a high concentration value of the molecule, either IC or NR depending on the transmitter, and by modulating bit-0 onto the absence of the molecule, thus injecting only the solvent PBS. During the experiments, $\rm TX_1$ was emitting the bit sequence ``1001000'' while $\rm TX_2$ was emitting the sequence ``1101001'', corresponding respectively to the ASCII characters ``H'' and ``i''. The syringe pumps at the transmitters were respectively loaded with an IC concentration of $2.18 \times 10^{-5}$~mol/L and a NR concentration of $1.15 \times 10^{-4}$~mol/L. With the dilutions occuring at the T-junction between the two transmitters and the receiver, the final concentrations of each molecule reaching the receiver should be respectively of {$1.09 \times 10^{-5}$} and {$5.73 \times 10^{-5}$~mol/L} only when the output flow rates of the two transmitters are the same.

\begin{figure}[!t]
	\centering
	\includegraphics[width=\linewidth]{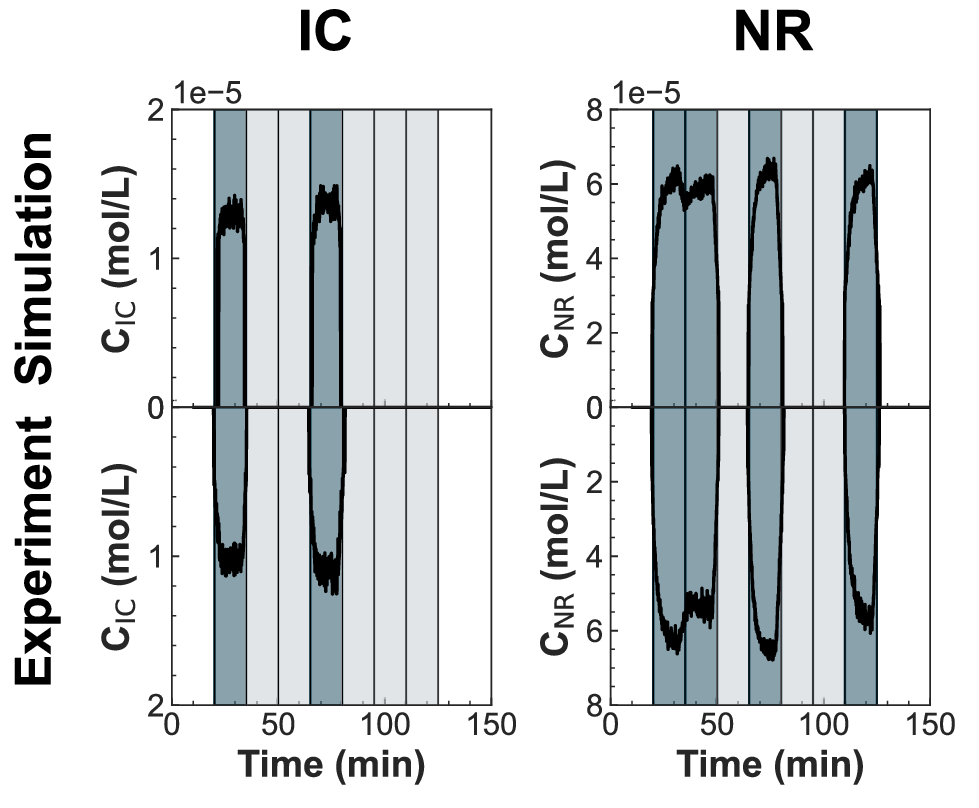}
	\caption{The detected evolution of the concentrations of IC and NR under synchronized BCSK transmission.}
	\label{f_decoded_sync}
\end{figure}
For the first experiment, the two transmitters used the same bit interval \todo{($T_b=15$ min)} and duty cycle \todo{($\alpha=1$)} for each bit and started to emit their respective bit sequences at the same time, meaning that the two transmitters were well synchronized. \todo{For each transmitter, when bit-1 is transmitted, the flow rate of information molecules is set to 40 $\mu$L/min, and the paired pump injecting PBS is turned off. Conversely, when bit-0 is transmitted, the information molecule pump is off, and the PBS pump operates at a flow rate of 40 $\mu$L/min.} The time-varying absorbance measured by the UV-Vis spectrometer at the receiver of the platform was analyzed by the two trained fCNN models (by experimental and simulated datasets) to detect the evolution of the concentration of each molecule over time.
Fig. \ref{f_decoded_sync} plots the detected bits obtained using both models and the comparison with the expected input bits indicated by the alternating background shading.
We clearly see that both models can perfectly detect the information emitted from both transmitters, with each bit accurately recovered. We also compare the average detected concentrations over a bit interval with the real experimental concentrations in Table~\ref{table:bcsk_sync}. It shows that the fCNN model trained on the simulated dataset with sensor noise always slightly overestimates the concentration by about $2.5 \times 10^{-6}$~mol/L for both molecules. However, this deviation remains within the range of the detection error measured during model validation (see Table~\ref{table:model_accuracy}).

\begin{comment}
    \begin{table}[!t]
    \caption{Comparison of the concentrations used to encode the bit-1 at each transmitter, respectively in Indigo Carmine (IC) for transmitter 1 and Neutral Red (NR) for transmitter 2, as really used and as predicted by either the experimental or the simulation fCNN model. \db{@Vivien: Is the decoded concentration an average value of the received signal within a bit interval?} {\color{red}Yes}}
    \label{table:bcsk_sync}
    \centering
    \begin{tabular}{|l||c|c|}
    \hline
     & \textbf{Transmitter 1} & \textbf{Transmitter 2} \\
    \hline
    Real & $1.09 \times 10^{-5}$ & $5.73 \times 10^{-5}$ \\
    \hline
    Experimental model & $1.06 \times 10^{-5}$ & $5.81 \times 10^{-5}$ \\
    \hline
    Simulation model & $1.33 \times 10^{-5}$ & $5.99 \times 10^{-5}$ \\
    \hline
    \end{tabular}
    \end{table}
\end{comment}

\begin{table}[!t]
\caption{Comparison between the actual concentrations and the detected concentrations obtained from the experimental- and simulation-trained fCNN models under synchronized BCSK transmission. All values are given in mol/L.}
\label{table:bcsk_sync}
\centering
\begin{tabular}{|l||c|c|}
\hline
\textbf{Model} & \textbf{Transmitter 1 (IC)} & \textbf{Transmitter 2 (NR)} \\
\hline
Real & $1.09 \times 10^{-5}$ & $5.73 \times 10^{-5}$ \\
\hline
Experimental fCNN & $1.06 \times 10^{-5}$ & $5.81 \times 10^{-5}$ \\
\hline
Simulation fCNN & $1.33 \times 10^{-5}$ & $5.99 \times 10^{-5}$ \\
\hline
\end{tabular}
\end{table}

\begin{figure}[!t]
	\centering
	\includegraphics[width=\linewidth]{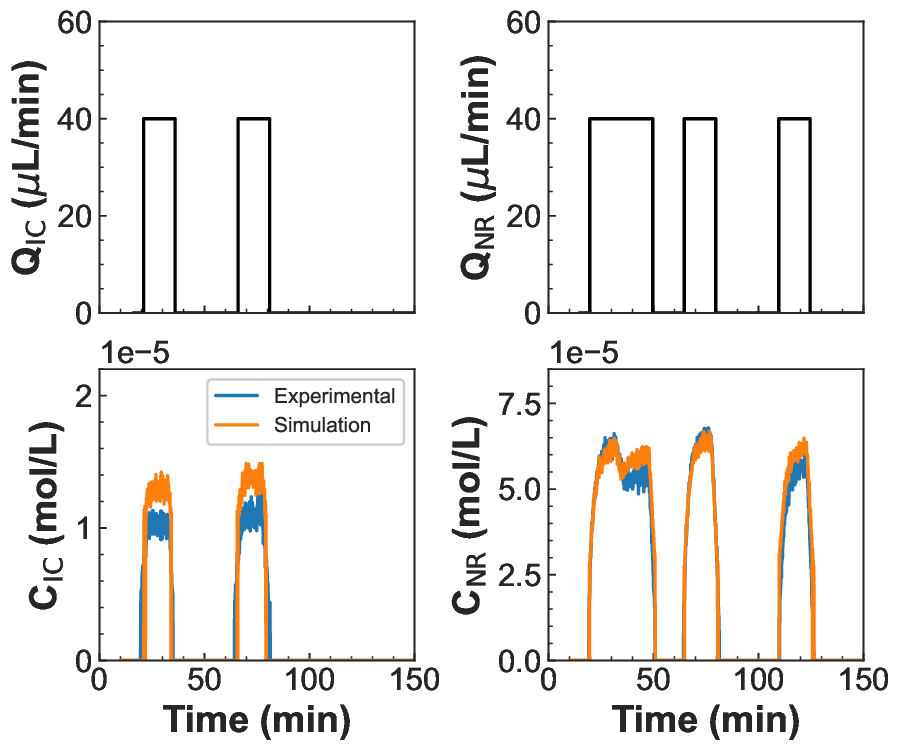}
	\caption{The flow rate profiles of IC and NR and the decoded bit sequences under synchronized BCSK.}
	\label{f_message_sync}
\end{figure}

Fig. \ref{f_message_sync} also compares the detected bit sequences produced by the two models along with the input syringe flow rates of the two transmitters. First, it further highlights the tendency of the model trained on the simulated dataset to overestimate concentration compared to the model trained on the experimental dataset, particularly for IC. Second, we can observe a slight variation in the evolution of the concentration of NR for $\rm TX_2$ during the emission of the second bit-1 of the sequence, corresponding to the moment that $\rm TX_1$ transitioned from emitting a bit-1 via IC to a bit-0 via PBS. At this point, both models exhibit a drop in the detection concentration of NR. For the model trained on the experimental dataset, the deviation reaches as high as $8.72 \times 10^{-6}$~mol/L, which exceeds its previously measured detection error. In contrast, the model trained on the simulated dataset with sensor noise detects a smaller variation of only $1.40 \times 10^{-6}$~mol/L, which falls within its detection error range.

%In a second experiment, the two transmitters were desynchronised: not only can they start emitting at any time, regardless of the other transmitter's status, but they can encode using different bit interval. In this case, the emission of a bit from a transmitter can finish while the other transmitter is still emitting a bit itself.

\begin{figure}[!t]
	\centering
	\includegraphics[width=\linewidth]{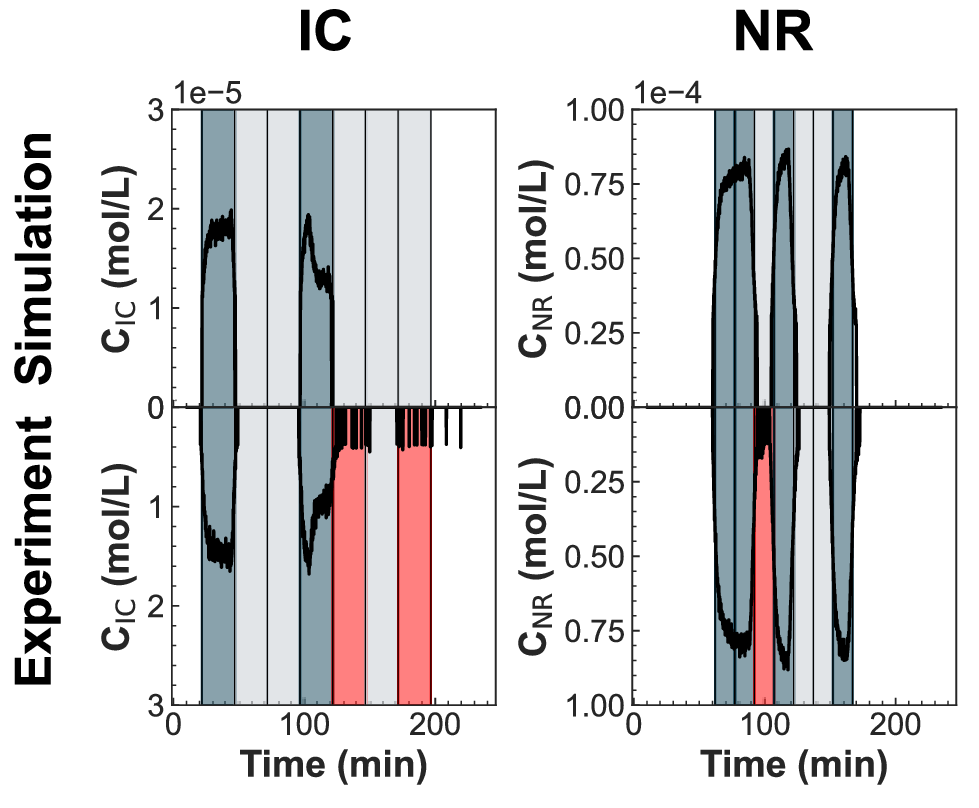}
	\caption{The detected evolution of the concentrations of IC and NR under desynchronized BCSK transmission.}
	\label{f_decoded_desync}
\end{figure}

For the second experiment, the two transmitters used different bit intervals \todo{($T_b=25$ min for $\rm TX_1$ and $T_b=15$ min for $\rm TX_2$)} and emitted signals at different times, meaning that the two transmitters were desynchronized. \todo{For $\rm TX_1$, when transmitting bit-1, the information chemical (IC) is injected at a constant flow rate of 25 $\mu$L/min with the paired PBS pump turned off; for bit-0, the configuration is reversed. For $\rm TX_2$, the NR flow rate is set to 40 $\mu$L/min under the same bit-dependent switching scheme. The duty cycle is maintained at $\alpha = 1$ for both transmitters.} Fig. \ref{f_decoded_desync} plots the comparison between the input message and the decoded message. We observe that both models can still successfully detect the transmitted bit-1s for both transmitters, but the model trained on the experimental dataset fails to decode bit-0 sometimes (highlighted in Salmon color). The superior performance of the simulation-trained model can be attributed to its ability to learn from an effectively unlimited number of concentration combinations, which is infeasible to replicate experimentally due to practical time constraints, e.g., generating only 30 IC-NR mixtures required an entire day of experimental effort.

\begin{figure}[!t]
	\centering
	\includegraphics[width=\linewidth]{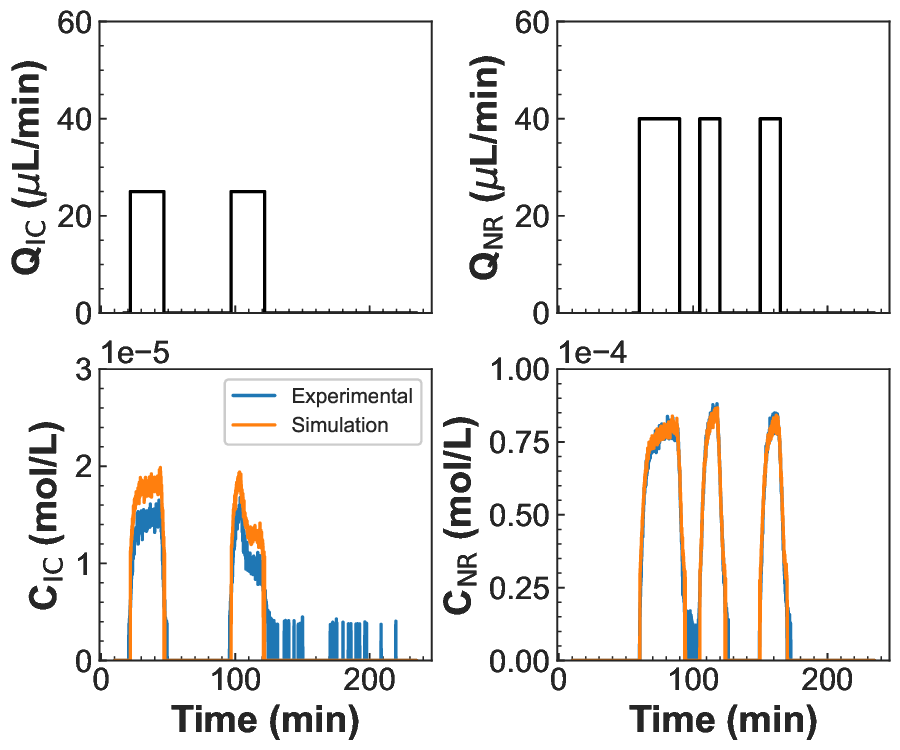}
	\caption{The flow rate profiles of IC and NR and the decoded bit sequences under desynchronized BCSK.}
	\label{f_message_desync}
\end{figure}

Similar to the synchronized case, Fig. \ref{f_message_desync} also compares the detected bit sequences produced by the two models along with the input syringe flow rates of the two transmitters. We can also observe that the model trained on the simulated dataset with sensor noise tends to overdetect the concentration of IC, and there is a IC concentration drop of the second bit-1 for $\rm TX_1$ as $\rm TX_2$ started to transmit bit-1 via NR.

\subsubsection{QCSK}
\todo{In QCSK, the signal was converted at each transmitter by encoding two bits into four concentration levels (i.e., $C_0$, $C_1$, $C_2$, and $C_3$) of the molecule. The mapping between the bit combination and the transmitted flow rate is summarized in Table \ref{table:qcsk_flowrate}. During experiments, $\rm TX_1$ was emitting the bit sequences ``1001011'' and ``1000011'' simultaneously, whereas $\rm TX_2$ was emitting the bit sequences ``1001100'' and ``0100001'' simultaneously. These corresponds to the letters ``K'' and ``C'' for $\rm TX_1$, and to the letters ``L'' and ``!'' for $\rm TX_2$, with the complete transmitted message ``KCL!''. The two transmitters were well synchronized with the same bit interval ($T_b=15$ min) and duty cycle ($\alpha=1$). 
}

\begin{figure}[!t]
	\centering
	\includegraphics[width=\linewidth]{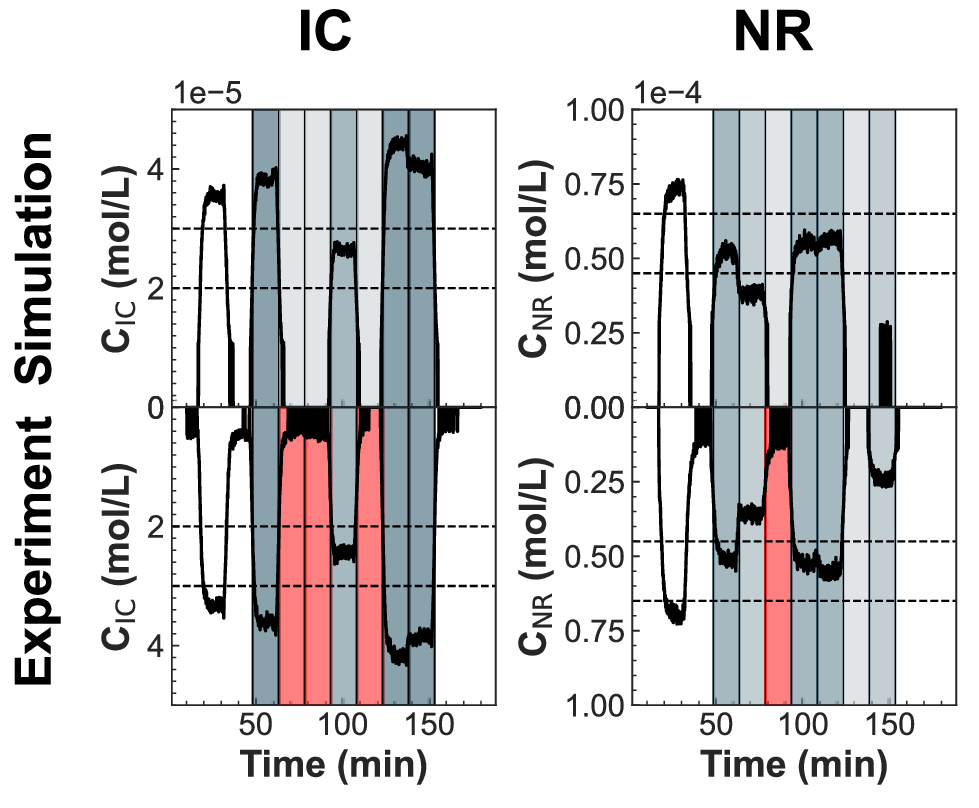}
	\caption{The detected evolution of the concentrations of IC and NR under synchronized QCSK transmission.}
	\label{f_decoded_qcsk}
\end{figure}
\begin{table}[!t]
    \centering
    \caption{Flow rate configuration for different levels and transmitters. All flow rate values are given in $\mu$L/min.}
        \begin{tabular}{|c|c||c|c|c|c|}
            \hline
            \multirow{2}{*}{\centering\textbf{Concentration}} & \multirow{2}{*}{\centering\textbf{Bit Combination}} 
            & \multicolumn{2}{c|}{\textbf{Transmitter 1}} & \multicolumn{2}{c|}{\textbf{Transmitter 2}} \\
            \cline{3-6}
            & & \textbf{IC} & \textbf{PBS} & \textbf{NR} & \textbf{PBS} \\
            \hline
            $C_0$ & 00 & 0 & 60 & 0 & 60 \\
            \hline
            $C_1$ & 01 & 20 & 40 & 20 & 40 \\
            \hline
            $C_2$ & 10 & 40 & 20 & 40 & 20 \\
            \hline
            $C_3$ & 11 & 60 & 0 & 60 & 0 \\
            \hline
        \end{tabular}
    \label{table:qcsk_flowrate}
\end{table}

\begin{table*}[!t]
    \centering
    \caption{Comparison between the actual concentrations and the detected concentrations obtained from the experimental- and simulation-trained fCNN models under synchronized QCSK transmission. All values are given in mol/L.}
        \begin{tabular}{|l||c|c|c|c|c|c|}
            \hline
            \multirow{2}{*}{\textbf{Model}} 
            & \multicolumn{3}{c|}{\textbf{Transmitter 1 (IC)}} 
            & \multicolumn{3}{c|}{\textbf{Transmitter 2 (NR)}} \\
            \cline{2-7}
            & $C_1$ & $C_2$ & $C_3$
            & $C_1$ & $C_2$ & $C_3$ \\
            \hline
            {Real} & $1.2\times10^{-5}$ & $2.4\times10^{-5}$ & $3.6\times10^{-5}$ & $2.9\times10^{-5}$ & $5.7\times10^{-5}$ & $8.6\times10^{-5}$ \\
            \hline
            {Experimental fCNN} & -- & $2.4\times10^{-5}$ & $2.6\times10^{-5}$ & $2.9\times10^{-5}$ & $5.1\times10^{-5}$ & -- \\
            \hline
            {Simulation fCNN} & -- & $2.6\times10^{-5}$ & $4.0\times10^{-5}$ & $3.8\times10^{-5}$ & $5.3\times10^{-5}$ & -- \\
            \hline
    \end{tabular}

    \vspace{1mm}
    \begin{minipage}{0.95\textwidth}
        \footnotesize \text{Note:} ‘--’ indicates that the corresponding concentration level was not used in transmitting the message ``KCL!''.
    \end{minipage}
    \label{table:qcsk_cons}
\end{table*}
Fig.~\ref{f_decoded_qcsk} illustrates the decoded bit sequences obtained using both the simulation-trained model and the experimental-trained model, overlaid with the transmitted message represented by shaded backgrounds—ranging from dark to light grey for increasing concentration levels (from 3 to 0). The comparison between the average detected concentrations over a bit interval with the real experimental concentrations is also shown in Table \ref{table:qcsk_cons}. We observe that the simulation-trained model can perfectly recover the message, achieving a bit error rate (BER) of 0\% over the 14 transmitted bits. In contrast, detecting with the experimentally trained model results in four bit errors (highlighted in Salmon color): three for IC and one for NR, yielding a BER of approximately 29\%. This further reveals the limitations inherent in the experimental dataset.

\begin{figure}[!t]
	\centering
	\includegraphics[width=0.65\linewidth]{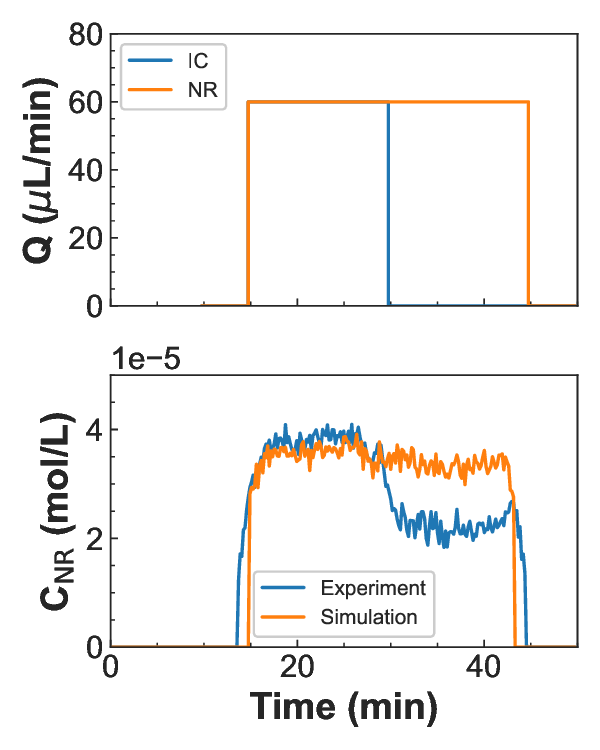}
	\caption{Illustration of the detection error exhibited by the model trained on the experimental dataset when the flow rate configuration changes.}
	\label{f_qcsk_error}
\end{figure}

\todo{
Fig.~\ref{f_qcsk_error} further reveals the underlying cause of the decoding errors observed in Fig.~\ref{f_decoded_qcsk}. Specifically, it highlights the sensitivity of the experimental model to changes in environmental conditions. When both IC and NR are emitted simultaneously at the same flow rate, and IC is subsequently replaced by PBS to maintain a constant total flow rate, the experimental model erroneously detects a drop in NR concentration—despite the NR flow remaining unchanged. 
%This artefactual fluctuation is not observed in the actual concentration, indicating a limitation in the model’s generalization ability. 
In contrast, the simulation-trained model correctly maintains a stable detection of the NR concentration under the same conditions, demonstrating greater robustness to flow configuration changes.
}

\section{Conclusion}
In this work, we proposed a fractal convolutional neural network to detect chemical signals in a multi-transmitter molecular communication platform. By leveraging UV-Vis spectroscopy, we demonstrated that our model can accurately detect chemical concentrations and transmitted messages, even in the presence of signal interference and temporal desynchronization. Importantly, we showed that simulated datasets—when augmented with realistic sensor noise—can serve as an effective substitute for experimental training data, significantly reducing the effort and cost associated with experimental data collection. Our findings open the door to scalable, data-driven molecular communication systems and pave the way for robust machine learning-based signal demodulation in future bio-hybrid networks.

\label{sec_conclusion}

% or

\appendices


%\section*{Acknowledgment}
%{\color{red}This work was supported by the US National Science Foundation (NSF) through grant MCB-1449014, and the NSF Nebraska EPSCoR through the First Award grant EPS-1004094.}


\ifCLASSOPTIONcaptionsoff
  \newpage
\fi



%\bibliographystyle{IEEEtran}
%\bibliography{Ref}

\begin{thebibliography}{10}
\providecommand{\url}[1]{#1}
\csname url@samestyle\endcsname
\providecommand{\newblock}{\relax}
\providecommand{\bibinfo}[2]{#2}
\providecommand{\BIBentrySTDinterwordspacing}{\spaceskip=0pt\relax}
\providecommand{\BIBentryALTinterwordstretchfactor}{4}
\providecommand{\BIBentryALTinterwordspacing}{\spaceskip=\fontdimen2\font plus
\BIBentryALTinterwordstretchfactor\fontdimen3\font minus \fontdimen4\font\relax}
\providecommand{\BIBforeignlanguage}[2]{{%
\expandafter\ifx\csname l@#1\endcsname\relax
\typeout{** WARNING: IEEEtran.bst: No hyphenation pattern has been}%
\typeout{** loaded for the language `#1'. Using the pattern for}%
\typeout{** the default language instead.}%
\else
\language=\csname l@#1\endcsname
\fi
#2}}
\providecommand{\BIBdecl}{\relax}
\BIBdecl

\bibitem{Akan2017}
O.~B. Akan, H.~Ramezani, T.~Khan, N.~A. Abbasi, and M.~Kuscu, ``Fundamentals of molecular information and communication science,'' \emph{Proc. IEEE}, vol. 105, no.~2, pp. 306--318, Feb. 2017.

\bibitem{bi2021survey}
D.~Bi, A.~Almpanis, A.~Noel, Y.~Deng, and R.~Schober, ``A survey of molecular communication in cell biology: Establishing a new hierarchy for interdisciplinary applications,'' \emph{IEEE Commun. Surveys Tuts.}, vol.~23, no.~3, pp. 1494--1545, Mar. 2021.

\bibitem{Pierobon2010}
M.~Pierobon and I.~Akyildiz, ``A physical end-to-end model for molecular communication in nanonetworks,'' \emph{IEEE J. Sel. Areas Commun.}, vol.~28, no.~4, pp. 602--611, May 2010.

\bibitem{6708551}
D.~Kilinc and O.~B. Akan, ``Receiver design for molecular communication,'' \emph{IEEE J. Sel. Areas Commun.}, vol.~31, no.~12, pp. 705--714, Dec. 2013.

\bibitem{6191345}
K.~V. Srinivas, A.~W. Eckford, and R.~S. Adve, ``Molecular communication in fluid media: The additive inverse gaussian noise channel,'' \emph{IEEE Trans. Inf. Theory}, vol.~58, no.~7, pp. 4678--4692, Jul. 2012.

\bibitem{6949028}
H.~Li, S.~M. Moser, and D.~Guo, ``Capacity of the memoryless additive inverse gaussian noise channel,'' \emph{IEEE J. Sel. Areas Commun.}, vol.~32, no.~12, pp. 2315--2329, Dec. 2014.

\bibitem{Jamali2019b}
V.~Jamali, A.~Ahmadzadeh, W.~Wicke, A.~Noel, and R.~Schober, ``Channel modeling for diffusive molecular communication: A tutorial review,'' \emph{Proc. IEEE}, vol. 107, no.~7, pp. 1256--1301, Jul. 2019.

\bibitem{10105650}
S.~Lotter, L.~Brand, V.~Jamali, M.~Schäfer, H.~M. Loos, H.~Unterweger, S.~Greiner, J.~Kirchner, C.~Alexiou, D.~Drummer, G.~Fischer, A.~Buettner, and R.~Schober, ``Experimental research in synthetic molecular communications – {Part I},'' \emph{IEEE Nanotechnol. Mag.}, vol.~17, no.~3, pp. 42--53, Jun. 2023.

\bibitem{lotter2023experimental}
S.~Lotter, L.~Brand, V.~Jamali, M.~Sch{\"a}fer, H.~M. Loos, H.~Unterweger, S.~Greiner, J.~Kirchner, C.~Alexiou, D.~Drummer \emph{et~al.}, ``Experimental research in synthetic molecular communications--{Part II},'' \emph{IEEE Nanotechnol. Mag.}, vol.~17, no.~3, pp. 54--65, Jun. 2023.

\bibitem{10443866}
M.~Hamidović, S.~Angerbauer, D.~Bi, Y.~Deng, T.~Tugcu, and W.~Haselmayr, ``Microfluidic systems for molecular communications: A review from theory to practice,'' \emph{IEEE Trans. Mol. Biol. Multi-Scale Commun.}, vol.~10, no.~1, pp. 147--163, Mar. 2024.

\bibitem{9247172}
A.~A. Boulogeorgos, S.~E. Trevlakis, S.~A. Tegos, V.~K. Papanikolaou, and G.~K. Karagiannidis, ``Machine learning in nano-scale biomedical engineering,'' \emph{IEEE Trans. Mol. Biol. Multi-Scale Commun.}, vol.~7, no.~1, pp. 10--39, Mar. 2021.

\bibitem{8277667}
H.~B. Yilmaz, C.~Lee, Y.~J. Cho, and C.-B. Chae, ``{A machine learning approach to model the received signal in molecular communications},'' in \emph{Proc. IEEE BlackSeaCom}, Jun. 2017, pp. 1--5.

\bibitem{lee2017machine}
C.~Lee, H.~B. Yilmaz, C.-B. Chae, N.~Farsad, and A.~Goldsmith, ``Machine learning based channel modeling for molecular mimo communications,'' in \emph{Proc. IEEE SPAWC}, Jul. 2017, pp. 1--5.

\bibitem{9120354}
J.~Li, W.~Zhang, X.~Bao, M.~Abbaszadeh, and W.~Guo, ``Inference in turbulent molecular information channels using support vector machine,'' \emph{IEEE Trans. Mol. Biol. Multi-Scale Commun.}, vol.~6, no.~1, pp. 25--35, Jul. 2020.

\bibitem{8964317}
O.~D. Kose, M.~C. Gursoy, M.~Saraclar, A.~E. Pusane, and T.~Tugcu, ``Machine learning-based silent entity localization using molecular diffusion,'' \emph{IEEE Commun. Lett.}, vol.~24, no.~4, pp. 807--810, Apr. 2020.

\bibitem{10748363}
J.~T. Gómez, B.~D. Unluturk, F.-L. Lau, J.~Simonjan, R.~Wendt, S.~Fischer, and F.~Dressler, ``{DNA-Based Nanonetwork for Abnormality Detection and Localization in the Human Body},'' \emph{IEEE Trans. Nanotechnol.}, vol.~23, no. Nov., pp. 794--808, 2024.

\bibitem{8491088}
X.~Qian and M.~Di~Renzo, ``{Receiver Design in Molecular Communications: An Approach Based on Artificial Neural Networks},'' in \emph{Proc. IEEE ISWCS}, Aug. 2018, pp. 1--5.

\bibitem{kim2023machine}
S.-J. Kim, P.~Singh, and S.-Y. Jung, ``{A machine learning-based concentration-encoded molecular communication system},'' \emph{Nano Commun. Netw.}, vol.~35, p. 100433, Mar. 2023.

\bibitem{10059136}
O.~T. Baydas, O.~Cetinkaya, and O.~B. Akan, ``{Estimation and Detection for Molecular MIMO Communications in the Internet of Bio-Nano Things},'' \emph{IEEE Trans. Mol. Biol. Multi-Scale Commun.}, vol.~9, no.~1, pp. 106--110, Mar. 2023.

\bibitem{10130469}
V.~Selis, D.~T. McGuiness, and A.~Marshall, ``{A Novel ML-Based Symbol Detection Pipeline for Molecular Communication},'' \emph{IEEE Trans. Mol. Biol. Multi-Scale Commun.}, vol.~9, no.~2, pp. 207--216, Jun. 2023.

\bibitem{10904174}
C.~Xiang, Y.~Zhang, Y.~Huang, W.~Tan, X.~Chen, and M.~Wen, ``{Hybrid Recurrent Neural Network for Signal-Dependent Noise Suppression in Molecular Communication},'' \emph{IEEE Trans. Mol. Biol. Multi-Scale Commun.}, pp. 1--1, Feb. 2025.

\bibitem{9148818}
B.-H. Koo, H.~J. Kim, J.-Y. Kwon, and C.-B. Chae, ``{Deep Learning-based Human Implantable Nano Molecular Communications},'' in \emph{Proc. IEEE ICC}, Jun. 2020, pp. 1--7.

\bibitem{8255058}
N.~Farsad, D.~Pan, and A.~Goldsmith, ``{A Novel Experimental Platform for In-Vessel Multi-Chemical Molecular Communications},'' in \emph{Proc. IEEE GLOBECOM}, Dec. 2017, pp. 1--6.

\bibitem{10443671}
F.~Calì, S.~Barreca, G.~Li-Destri, A.~Torrisi, A.~Licciardello, and N.~Tuccitto, ``{Experimental Implementation of Molecule Shift Keying for Enhanced Molecular Communication},'' \emph{IEEE Trans. Mol. Biol. Multi-Scale Commun.}, vol.~10, no.~1, pp. 175--184, Feb. 2024.

\bibitem{bartunik2022using}
M.~Bartunik, O.~Keszocze, B.~Schiller, and J.~Kirchner, ``{Using deep learning to demodulate transmissions in molecular communication},'' in \emph{Proc. IEEE ISMICT}, May 2022, pp. 1--6.

\bibitem{deng2024channel}
K.~Deng, H.~Luo, W.~Zhang, Y.~Qin, J.~Zhang, Y.~Yuan, Y.~Wei, and L.~Lin, ``{Channel Parameter Estimation of Neural Communication Based on Deep Learning},'' in \emph{Proc. IEEE GLOBECOM}, Dec. 2024, pp. 2096--2101.

\bibitem{walter2023real}
V.~Walter, D.~Bi, A.~Salehi-Reyhani, and Y.~Deng, ``{Real-time signal processing via chemical reactions for a microfluidic molecular communication system},'' \emph{Nat. Commun.}, vol.~14, no.~1, p. 7188, 2023.

\bibitem{walters1997spectrophotometric}
C.~Walters, A.~Keeney, C.~T. Wigal, C.~R. Johnston, and R.~D. Cornelius, ``{The spectrophotometric analysis and modeling of sunscreens},'' \emph{J. Chem. Educ.}, vol.~74, no.~1, p.~99, Jan. 1997.

\bibitem{langergraber2003multivariate}
G.~Langergraber, N.~Fleischmann, and F.~Hofstaedter, ``{A multivariate calibration procedure for UV/VIS spectrometric quantification of organic matter and nitrate in wastewater},'' \emph{Water Sci. Technol.}, vol.~47, no.~2, pp. 63--71, Jan. 2003.

\bibitem{guan2018research}
L.~Guan, Y.~Tong, J.~Li, D.~Li, and S.~Wu, ``{Research on ultraviolet-visible absorption spectrum preprocessing for water quality contamination detection},'' \emph{Optik}, vol. 164, pp. 277--288, Mar. 2018.

\bibitem{lu2021detection}
Y.~Lu, X.~Li, W.~Li, T.~Shen, Z.~He, M.~Zhang, H.~Zhang, Y.~Sun, and F.~Liu, ``{Detection of chlorpyrifos and carbendazim residues in the cabbage using visible/near-infrared spectroscopy combined with chemometrics},'' \emph{Spectrochim. Acta A Mol. Biomol. Spectrosc.}, vol. 257, p. 119759, Apr. 2021.

\bibitem{wolf2013predicting}
C.~Wolf, D.~Gaida, A.~Stuhlsatz, T.~Ludwig, S.~McLoone, and M.~Bongards, ``{Predicting organic acid concentration from UV/vis spectrometry measurements--a comparison of machine learning techniques},'' \emph{Trans. Inst. Meas. Control.}, vol.~35, no.~1, pp. 5--15, Feb 2013.

\bibitem{larsson_2017}
G.~Larsson, M.~Maire, and G.~Shakhnarovich, ``{Fractalnet: Ultra-deep neural networks without residuals},'' \emph{arXiv:1605.07648}, May 2017.

\bibitem{xia_2023}
M.~Xia, R.~Yang, G.~Yin, X.~Chen, J.~Chen, and N.~Zhao, ``{A method based on a one-dimensional convolutional neural network for UV-vis spectrometric quantification of nitrate and COD in water under random turbidity disturbance scenario},'' \emph{RSC Adv.}, vol.~13, no.~1, pp. 516--526, 2023.

\bibitem{sabnis_book}
H.~O. A.~B. Indicators, \emph{Handbook Of Acid Base Indicators}.\hskip 1em plus 0.5em minus 0.4em\relax Florida, USA: CRC Press, 2007.

\end{thebibliography}

\end{document}